# Doppler-free intermodulated fluorescence spectroscopy of $^4$He $2^3$P-$3^{1,3}$D transitions at 588 nm with a one-watt compact laser system

Pei-Ling Luo · Jinmeng Hu · Yan Feng · Li-Bang Wang · Jow-Tsong Shy

**Abstract:** We have demonstrated Doppler-free intermodulated fluorescence spectroscopy of helium $2^3$P-$3^{1,3}$D transitions in an rf discharged sealed-off cell using a compact laser system at 588 nm. An external cavity diode laser at 1176 nm was constructed to seed a Raman fiber amplifier. Laser power of more than one watt at 588 nm was produced by frequency doubling of the fiber amplifier output using a MgO:PPLN crystal. A doubling efficiency of 23 % was achieved. The power-dependent spectra of the $2^3$P-$3^3$D transitions were investigated. Furthermore, the Doppler-free spectrum of the spin-forbidden $2^3$P-$3^1$D transitions was observed for the first time. Our results are crucial towards precision test of QED atomic calculations, especially for improving the determination of the helium $3^1$D-$3^3$D separation.

Pei-Ling Luo · Li-Bang Wang · Jow-Tsong Shy (✉)
Department of Physics and Frontier Research Center on Fundamental and Applied Sciences of Matters, National Tsing Hua University, Hsinchu, 30013 Taiwan
e-mail: shy@phys.nthu.edu.tw

Li-Bang Wang
e-mail: lbwang@phys.nthu.edu.tw

Jow-Tsong Shy
Institute of Photonics Technologies, National Tsing Hua University, Hsinchu, 30013 Taiwan

Jinmeng Hu · Yan Feng
Shanghai Key Laboratory of Solid State Laser and Application and Shanghai Institute of Optics and Fine Mechanics, Chinese Academy of Sciences, Shanghai 201800, China

## 1. Introduction

As the simplest multi-electron atom, helium is of both experimental and theoretical interest due to its calculable electronic structures. Precision spectroscopic measurements on the helium transitions, such as absolute transition frequencies [1-6], fine-structure intervals [7-11], and isotope shifts [1, 4, 12-13] have played essential roles for testing quantum electrodynamics (QED) in many-body atomic calculations [14-15]. The accuracy of the spectroscopic measurement depends on the linewidth and the signal to noise ratio (SNR) of the measured spectrum. Therefore, Doppler-free techniques have been widely used in precision spectroscopy [4-8]. Particularly, zero-background Doppler-free scheme, such as polarization spectroscopy [16, 17], intermodulated optogalvanic spectroscopy [18, 19], intermodulated fluorescence spectroscopy [20], resonance ionization spectroscopy [21] and crossed laser-atomic beam methods [5-8, 11, 12] are powerful techniques to improve spectral SNR for observing weak transitions.

The Doppler-free spectroscopy of the helium $2^3P$-$3^{1,3}D$ transitions is essential to determine the ionization energy of the $2^3P$ state and investigate the fine-structure and hyperfine intervals of the $2^3P$ and $3^3D$ states [18, 20]. Furthermore, it is also of interest to study the singlet-triplet mixing of the $3^1D$ and $3^3D$ states [20, 22]. Moreover, the separation between $3^3D$ and $3^1D$ states can be precisely determined by measuring the absolute frequencies of the Doppler-free spectra of $2^3P$-$3^{1,3}D$ transitions to resolve a discrepancy of 7.4 σ (37 MHz) between the theories and experiments of the $3^1D$-$3^3D$ separation in $^3$He [23]. The first observation of the Doppler-free spectrum of the $2^3P$-$3^3D$ transitions has been demonstrated by means of intermodulated optogalvanic spectroscopy in a dc discharge cell. The hyperfine intervals of the $^3$He $3^3D$ state were reported [18]. In a later report [20], Doppler-free spectra of the $^4$He $2^3P$-$3^3D$ transitions were performed in a dc discharged cell using intermodulated fluorescence spectroscopy. However, the Doppler-free spectrum of the spin-forbidden $2^3P$-$3^1D$ transitions could not be obtained. Only the Doppler-limited fluorescence spectrum of the $2^3P$-$3^1D$ transitions was observed by focusing 120 mW laser beam into the dc discharged cell. Up to now, no Doppler-free spectrum of the $2^3P$-$3^1D$ spin-forbidden transitions has been reported.

In these previous experiments, the spectroscopy of the helium $2^3P$-$3^{1,3}D$ transitions at 588 nm was performed using a dye laser system [18, 20, 22]. The laser power of a few hundred mW can be typically obtained. Nevertheless, the dye laser system is bulky and difficult to maintain. Thanks to the recent development of the quantum-dot semiconductor laser [24] and the nonlinear optical technique [25], a high power yellow laser based on frequency doubling of a diode laser system with a Raman fiber amplifier has been demonstrated [26], which is much more compact and stable than the dye laser.

In this letter, we report the observation on the Doppler-free intermodulated fluorescence spectroscopy of the $2^3P$-$3^{1,3}D$ transitions in $^4$He. The spectra were performed in an rf discharged cell using a compact laser system. Laser power of more than one watt at 588 nm was produced based on the efficient single-pass second-harmonic generation (SHG) of a fiber-coupled external cavity diode laser with a Raman fiber amplifier at 1176 nm. The Doppler-free signals of $2^3P$-$3^3D$ transitions in different laser power were studied. Most importantly, to the best of our knowledge, it was the first time that the Doppler-free spectroscopy of the spin-forbidden $2^3P$-$3^1D$ transitions was observed.

## 2. Experimental methods

A schematic drawing of the related helium energy levels is shown in Fig. 1(a). The helium atoms at the $2^3P$ states are produced by the radio-frequency (rf) discharge in a pure helium cell. For the measurement of the $2^3P$-$3^3D$ transitions, the population in the $2^3P$ states is excited to the $3^3D$ and the laser induced fluorescence of the $3^3D$-$2^3P$ transitions at 587.7 nm is detected. In the experiment of observation on the $2^3P$-$3^1D$ transition spectrum, the population in the $2^3P$ states is excited to $3^1D$ states and fluorescence of the $3^1D_2$-$2^1P_1$ transition at 668.0 nm is detected.

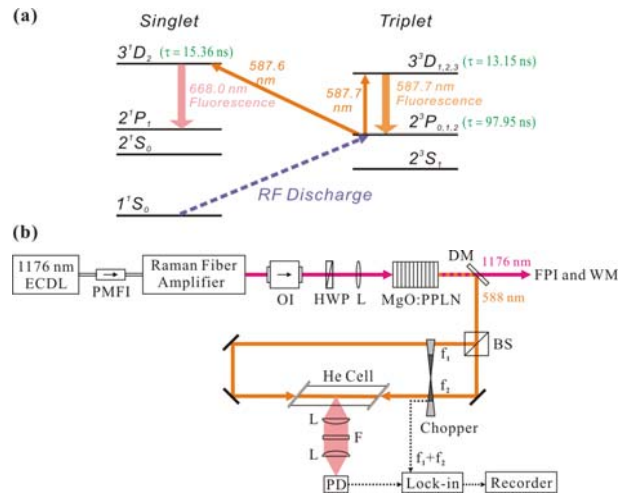

Fig. 1. (a) Energy level diagram for Doppler-free spectroscopy of the $^4$He $2^3P$-$3^{1,3}D$ transitions. (b) Schematic diagram of the experimental setup. ECDL, external cavity diode laser; PMFI, polarization-maintaining fiber isolator; OI, optical isolator; HWP, half-wave plate; L, lens; PPLN, periodically poled LiNbO$_3$ crystal; DM, dichroic mirror; FPI, Fabry-Perot interferometer; WM, wavemeter; BS, beam splitter; F, filter; PD, photodiode.

The experimental setup is shown in Fig. 1(b). The external cavity diode laser (ECDL) at 1176 nm was constructed in a polarization-maintaining (PM) fiber-coupled gain module (Innolume, GM-1140-120-PM-130) with the Littrow configuration. The laser wavelength was selected using a 1200 grooves/mm blazed grating. The grating was mounted on a piezoelectric transducer (PZT) to fine tune the angle of the grating. The wavelength of the ECDL can be tuned from 1100 to 1200 nm. Besides, the mode-hop-free tuning range over 10 GHz and the power output of a few tens mW were achieved. The output of the ECDL through two PM fiber isolators (PMFI) was sent to a stimulated Brillouin scattering (SBS) suppressed PM Raman fiber amplifier [13]. About 5 W of linearly-polarized narrow-linewidth laser power at 1176 nm was produced. An optical isolator (OI) at the output of the Raman fiber amplifier was used to avoid the optical feedback. The power amplified laser beam was focused into an MgO-doped periodically poled LiNbO$_3$ crystal (PPLN) for frequency doubling to 588 nm. The MgO:PPLN (HC Photonics) with a length of 50 mm and a period of 9.27 μm was temperature controlled at 95°C for 587.6 nm and 97°C for 587.7 nm to achieve the quasi-phase matching (QPM). The fundamental beam waist radius at the center of the MgO:PPLN was 50 μm which corresponds to a focusing parameter of 1.73. A dichroic mirror (DM) was placed in the output of the MgO:PPLN to separate the fundamental (1176 nm) and second-harmonic (588 nm) beams. The transmitted 1176 nm beam was sent to a Fabry-Perot interferometer (FPI) and a wavemeter (WM) for monitoring the laser operation condition. The reflective 588 nm beam was used to perform the spectroscopy of the $2^3P$-$3^{1,3}D$ transitions in atomic helium. Figure 2 shows the dependence of the power and efficiency of the SHG on the fundamental power. The 588 nm laser power of 1.1 W was generated at a fundamental power of 4.8 W. The frequency doubling conversion efficiency of 23 % was obtained.

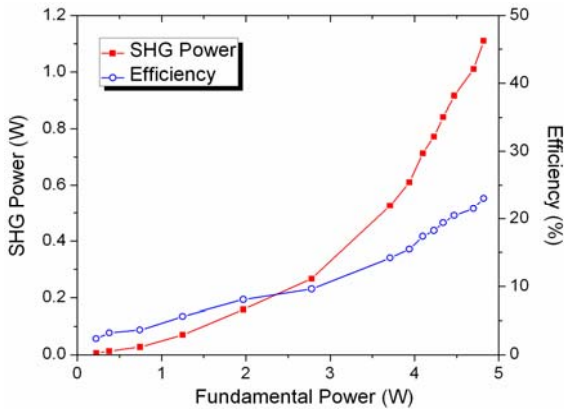

Fig. 2. The dependence of the SHG power and the efficiency on the fundamental power.

The measurement of the Doppler-free spectroscopy was performed by intermodulated fluorescence technique [27]. This technique utilizes the nonlinear effect between the two counter-propagating laser beams, which are resonant only with zero-velocity atoms. As a result, Doppler-free spectrum can be obtained. The 588 nm laser was separated into two beams by a 50:50 beam splitter (BS). Both beams were vertically polarized. One beam was chopped at 559 Hz and sent to a 15-cm rf discharged cell from one direction while another beam was chopped at 399 Hz and passed through the cell in the opposite direction. The sealed-off cell with Brewster windows was filled with pure $^4$He gas at 200 mTorr and was enclosed in three-layer mu-metal box for shielding the earth magnetic field to <1 mG. The laser induced fluorescence from the cell passing through a narrow bandpass filter was detected by a 9.8 mm-diameter photodetector (PD). The signal from PD was demodulated at the sum frequency of 958 Hz using a lock-in amplifier.

## 3. Results and discussion

Figure 3 shows power-dependent Doppler-free spectra of the $^4$He $2^3P$-$3^3D$ transitions. The two beams were overlapped in the cell with beam diameters of approximately 3.5 mm. The laser induced fluorescence of the $3^3D$-$2^3P$ transitions at 587.7 nm was detected. The Doppler-free spectra at six different total laser powers were presented. The power broadening spectra were clearly observed. Figure 4 shows the dependence of the full width at half maximum (FWHM) of the $^4$He $2^3P_1$-$3^3D_2$ ((a) in Fig. 3) on the light intensity. The FWHM at zero light intensity was determined to be 33(5) MHz which is comparable to the natural linewidth of 12 MHz with a pressure broadening of approximately 10 MHz at 200 mTorr [28]. The saturation intensity of the $2^3P_1$-$3^3D_2$ transition was determined to be 1.3(4)×10$^3$ W/m$^2$ which is in reasonable agreement with the calculated saturation intensity of 2.0×10$^3$ W/m$^2$, given by [29],

$$I_S = \frac{4\pi^2 hc\delta\nu}{\tau A \lambda^3} \qquad (1)$$

where $\delta\nu$ is the measured half width at half maximum (HWHM) at zero light intensity, $\tau^{-1} = \tau_1^{-1} + \tau_2^{-1}$, where $\tau_1$ and $\tau_2$ are the lifetime of the upper level and lower level respectively, $A$ is the Einstein $A$-coefficient of the transition, and $\lambda$ is the center wavelength of the transition. Furthermore, the signal of the $2^3P_2$-$3^3D_1$ transition ((h) in Fig. 3) was observed when the laser power was higher than 180 mW. In previous experiments [30], the $2^3P_2$-$3^3D_1$ transition was too weak to be observed since the line strength of $2^3P_2$-$3^3D_1$ transition is 45 times weaker than the $2^3P_1$-$3^3D_2$ transition and the saturation intensity of the $2^3P_2$-$3^3D_1$ transition is 27 times the value of the $2^3P_1$-$3^3D_2$ transition. The whole power-dependent spectra of the $2^3P$-$3^3D$ transition lines in $^4$He were demonstrated. Our results are significant and will be the first mile stone towards measuring absolute transition frequencies and fine-structure intervals to test QED atomic calculations.

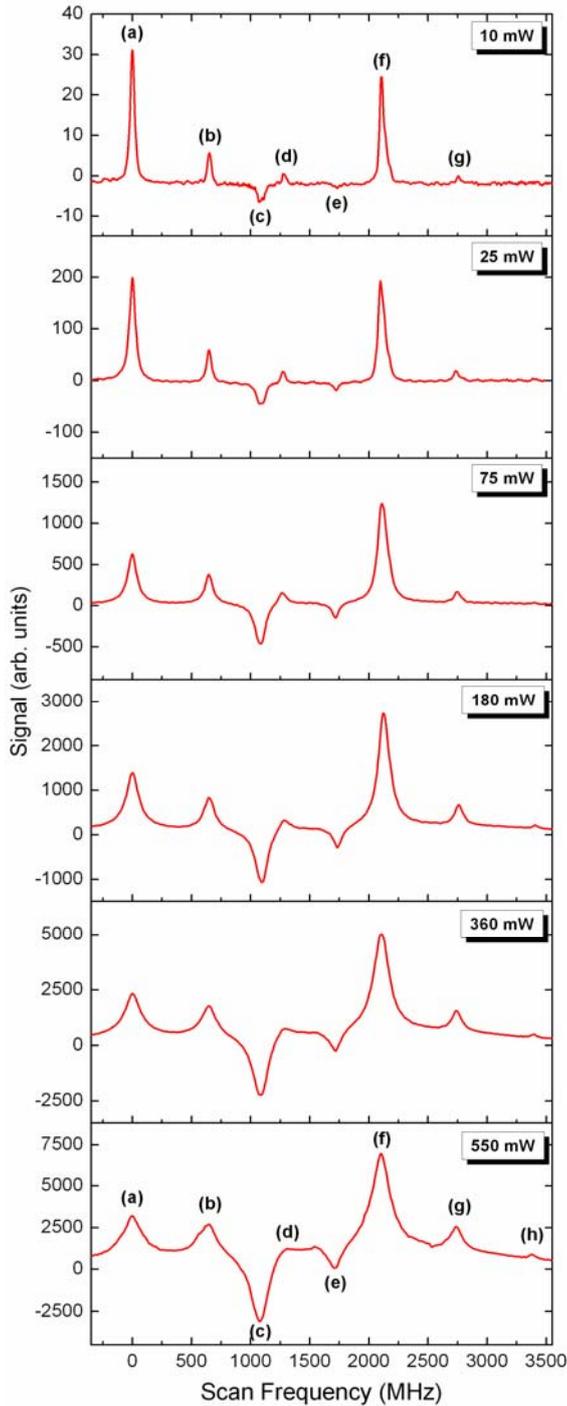

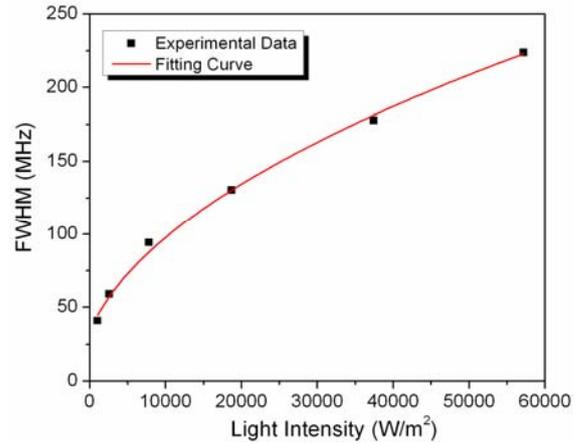

Fig. 4. The dependence of the full width at half maximum (FWHM) of the $^4$He $2^3P_1$-$3^3D_2$ on the light intensity. The cell pressure is 200 mTorr and the rf discharge power is 22 W.

Fig. 3. The power-dependent Doppler-free spectra of the $2^3P$-$3^3D$ transitions in $^4$He. The transitions are (a) $2^3P_1$-$3^3D_2$, (d) $2^3P_1$-$3^3D_1$, (f) $2^3P_2$-$3^3D_3$ and $2^3P_2$-$3^3D_2$, and (h) $2^3P_2$-$3^3D_1$. The cross-over lines with the common lower levels are (b) and (g). The cross-over lines with the common upper levels are (c) and (e). Here, the cell pressure is 200 mTorr. The rf discharge power is 22 W. The laser power is the total power of both beams and the beam diameter (1/e) is approximately 3.5 mm.

Most importantly, the Doppler-free spectra of the $^4$He spin-forbidden $2^3P_{1,2}$-$3^1D_2$ transitions as well as the cross-over line were obtained, as shown in Fig. 5. Before our measurement, only Doppler-limited spectra of the spin-forbidden $2^3P_{1,2}$-$3^1D_2$ transitions of $^4$He were observed [20]. The main difficulty is that the line strength of the $2^3P_1$-$3^1D_2$ and $2^3P_2$-$3^1D_2$ transitions are $0.4\times10^4$ and $1.2\times10^4$ times weaker than the $2^3P_1$-$3^3D_2$ transition. In addition, the estimated saturation intensity of the $2^3P_{1,2}$-$3^1D_2$ transitions is three to four orders of magnitude higher than that of the $2^3P_1$-$3^3D_2$ transition. To obtain the saturated-absorption signal, we focused the laser beam to a diameter of approximately 0.4 mm in the center of the cell to achieve enough light intensity. The laser induced fluorescence of the $3^1D$-$2^1P$ transitions at 668.0 nm was detected as the signal. Compared with the spectra of the $2^3P$-$3^3D$ transitions, the spectra of the $2^3P_{1,2}$-$3^1D_2$ transitions show asymmetric lineshapes. The mechanism for producing this peculiar lineshape is still under investigation. However, we have found that by changing the relative polarization of the two laser beams, the asymmetry can be reduced. In the future, we will use an optical frequency comb stabilized laser system [31] to study the origin of this asymmetric lineshape. Optical frequency measurements of both $2^3P$-$3^3D$ and $2^3P$-$3^1D$ transitions will be performed to determine the $3^1D$-$3^3D$ separation as well as the singlet-triplet mixing coefficient. A measurement on $^3$He will also help to resolve the discrepancy between theory and former measurements. The anticipated error budget of these measurements is listed in Table 1. As we fit the line profile for a single scan, the uncertainty in determining the line center is approximately 100 kHz and 1 MHz for $2^3P$-$3^3D$ and $2^3P$-$3^1D$ transitions respectively. The statistical uncertainty can then be reduced for 20 or more scans. The uncertainty of pressure shift in $2^3P$-$3^1D$ transition is expected to be much larger because the spin-forbidden signal can only be observed with larger helium pressure (> 200 mTorr) and therefore systematic study of the pressure shift can only be performed for the $2^3P$-$3^3D$ transition. The effect due to the asymmetric lineshape is difficult to estimate. In our analysis, we decompose the spectrum into a Lorenzian part and a dispersive part and find the line center of the Lorenzian part does not vary at different experimental conditions. Here we conservatively quote 2 MHz for this effect and can be further reduced once the mechanism is studied and understood.

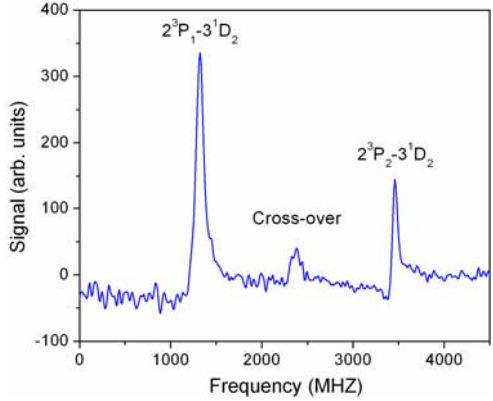

Fig. 5. The Doppler-free spectra of the $^4$He spin-forbidden $2^3P$-$3^1D$ transitions and their cross-over line. Here, the cell pressure is 200 mTorr. The rf discharge power is 22 W. The beam diameter is approximately 0.4 mm and the total laser power is 1 W.

Table 1. The anticipated uncertainties in the determined absolute transition frequencies.

| Sources | $2^3P$-$3^3D$ | $2^3P$-$3^1D$ |
| --- | --- | --- |
| Statistical | 30 kHz | 300 kHz |
| Frequency comb accuracy | 2 kHz | 2 kHz |
| Frequency locking stability | 2 kHz | 2 kHz |
| Zeeman shift (B field < 1 mG) | < 1 kHz | < 1 kHz |
| Pressure shift | 10 kHz | 200 kHz |
| Asymmetric lineshape |  | 2 MHz |
| Total | 44 kHz | 2.3 MHz |

In conclusion, we have constructed a compact laser system at 588 nm based on highly-efficient SHG of a 1176 nm ECDL-seeded Raman fiber amplifier. The laser has an optical power of over 1 W at 588 nm and it can be easily controlled and maintained. With this compact laser system, the Doppler-free spectra of the whole $2^3P$-$3^{1,3}D$ transition lines in $^4$He were observed using intermodulated fluorescence detection scheme. The Doppler-free signals of the weak $2^3P_2$-$3^3D_1$ transition and the spin-forbidden $2^3P_{1,2}$-$3^1D_2$ intercombination transitions were successfully observed for the first time. Further measurements on absolute transition frequencies, fine structure intervals, and isotope shifts of helium $2^3P$-$3^{1,3}D$ transitions are essential to test QED atomic calculations. Most importantly, accurate determination of the $3^1D$-$3^3D$ separation in $^3$He is critical to resolve the 7.4 σ discrepancy between the theories and experiments [23]. In addition, as suggested in [32], this intercombination transition at 588 nm can be used for preparing ultracold helium atoms in the $2^1S$ state, and this is the first step towards probing $1^1S$-$2^1S$ two-photon transition by direct frequency comb spectroscopy in a magneto-optical trap. Our result clearly has demonstrated successful transfer of the atomic populations from the triplet states to the singlet states.

**Acknowledgment**

This project is supported by the Ministry of Science and Technology and the Ministry of Education of Taiwan. L.-B.W. receives support from Kenda Foundation as a Golden-Jade fellow.